\documentstyle[11pt,newpasp,twoside,psfig,wrapfig]{article}
\index{3EG~J2227+6122,G106.3+2.7,G106.6+2.9,PSR B0833--45,PSR~J2229+6114,Vela Pulsar}
\index{instructions}
\index{template}

\def\source{3EG J2227+6122}

\def\asca{{\it ASCA\/}}
\def\chandra{{\it Chandra\/}}

\def\psr{PSR J2229+6114}
\def\snr{G106.6+2.9}

\def\edcomment#1{\iffalse\marginpar{\raggedright\sl#1\/}\else\relax\fi}
\reversemarginpar

\begin{document}
\title{The New Gamma-ray Pulsar PSR J2229+6114, its Pulsar Wind Nebula,
and Comparison with the Vela Pulsar}
\author{J. P. Halpern, E. V. Gotthelf, F. Camilo, B. Collins, \& D. J. Helfand}
\affil{Columbia University, 550 West 120th Street, New York, NY 10027}

\begin{abstract}
With a period of 51.6 ms and spin-down power
$2.2 \times 10^{37}$ erg~s$^{-1}$, \psr\
is a compelling identification for the EGRET source \source\ in which
error circle it resides.  Striking features of
the \chandra\ X-ray image are an incomplete 
elliptical arc and a possible jet,
similar to the structures that dominate the appearance of the Vela PWN.
Approximately 70\% of the 2--10 keV X-ray emission comes
from a centrally peaked, diffuse nebula of radius $100^{\prime\prime}$
with a power-law spectrum of photon index $1.45 \pm 0.19$.
The pulsar itself has a marginally harder spectrum with photon
index $0.99 \pm 0.27$.
For an assumed distance of 3 kpc the ratio of X-ray 
luminosity to spin-down power of \psr\ is only $8 \times 10^{-5}$.
We discuss a model in which such inefficient X-ray emission is 
the signature of a highly magnetized pulsar wind that prevents an internal MHD 
shock, at the location of the X-ray arc, from strongly compressing the flow. 
The incomplete X-ray arc is consistent with beaming from a relativistic 
equatorial outflow, while a surrounding radio shell is probably
the forward shock in the surrounding ISM.

An MeV source at this location was previously detected
by the COMPTEL experiment on {\it CGRO\/}.
This, plus the flat X-ray spectrum in the 
2--10 keV band, suggests that \psr\ is one of the brightest pulsars
at 1 MeV, even while it is inconspicuous at radio through X-ray wavelengths,
and as steep as the Crab above 100 MeV.  The apparent variety of broad-band 
spectra displayed by high-energy pulsars bolsters the theory that
rotation-powered pulsars
dominate the unidentified Galactic EGRET
source population.  
\end{abstract}

\section{Does PSR J2229+6114 = 3EG J2227+6122?}

\source\ (Hartman et al. 1999)
is one of many ``unidentified'' EGRET sources at low
Galactic latitude, $(\ell,b) = (106.\!^{\circ}5,3.\!^{\circ}2)$, for
which a pulsar origin is the hypothesis favored by many
authors.  Halpern et al. (2001a,b) made a complete
multiwavelength study of the error circle of \source\
that culminated in the discovery of a single plausible
counterpart, a young, energetic pulsar
(Figures 1 and 2) with an associated X-ray PWN
(Figure 3) that is confined within a small, nonthermal radio shell.
Compared to known $\gamma$-ray pulsars \psr\ is
second only to the Crab in spin-down power, and it is significantly more
luminous than the Vela pulsar (PSR B0833--45).
For a distance of 3~kpc estimated from X-ray absorption, 
\psr\ ranks \#3 or \#4 among all pulsars in spin-down flux $\dot E/d^2$.
If \psr\ is the counterpart of \source, then
its luminosity above 100~MeV is $\approx 3.7 \times 10^{35}$
erg~s$^{-1}$, and its efficiency $\eta$ of $>100$~MeV
$\gamma$-ray production, if isotropic, is $0.016\,(d/3\,{\rm kpc})^2$.
Among the pulsars that are either reliably
or probably identified with
EGRET sources (Table 1), there is a trend (e.g., Thompson et al. 1999)
in which the efficiency of $\gamma$-ray production increases with
decreasing spin-down power ($\dot E \propto B^2/P^4$), or equivalently,
open field line voltage ($\Phi \propto B/P^2$).
As the source of \source, \psr\ would have an efficiency in accord 
with the established pattern if its distance were close to our
estimate of 3~kpc.
It's a pretty good bet that \psr\ is responsible for \source\
even though the EGRET photons are too few and too old for a significant,
confirming pulsar detection using the contemporary ephemeris
(Thompson et al., these proceedings).
If \psr\ is not the correct identification, it implies that a highly efficient
(or highly beamed) $\gamma$-ray source can avoid producing 
soft or hard X-rays, in this case at a level below $4 \times 10^{-14}$
erg cm$^{-2}$ s$^{-1}$, or $< 10^{-4}$ of its apparent
$\gamma$-ray luminosity, which is unprecedented.
Taking into account our exhaustive search of the error circle, the
conservative conclusion is acceptance of
the identification of \source\ with \psr.

\begin{figure}
\psfig{figure=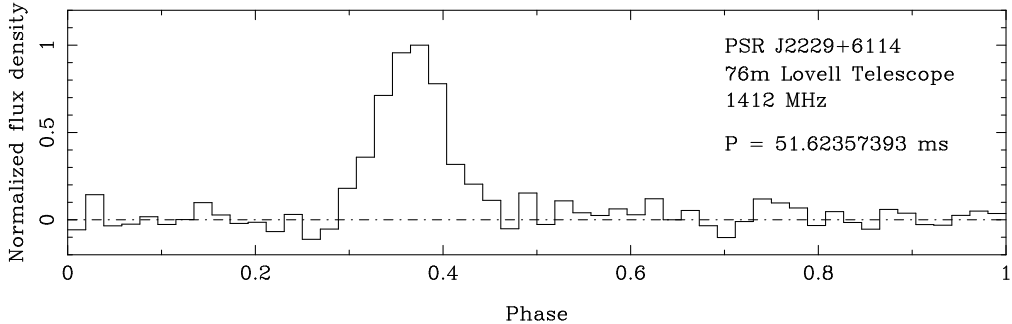,width=5.25truein,angle=270.}
\caption{Radio pulse profile of \psr\ at 1412~MHz. 
The instrumental resolution is $\approx 0.02$ of the period.
Phase zero is arbitrary.
The period-averaged flux density is only $\approx 0.25$~mJy,
consistent with its non-detection in previous ``all-sky'' 
pulsar surveys.}
\end{figure}

\begin{table}
\caption{EGRET PULSARS}
\begin{tabular}{lcrrcc}
\tableline
\hfil Name \hfil & Period & Age $(P/2\dot P)$ & Dist. &
$\dot E = I \Omega \dot \Omega$ & $\eta$ \\
& (s) & \hfil (yr) \hfil & \hfil (pc) \hfil & (erg s$^{-1}$) & ($>100$ MeV) \\
\tableline
Crab           & 0.033 & 1,250   & 2,000 & $5.0 \times 10^{38}$ & 0.002    \\
PSR J2229+6114 & 0.051 & 10,500  & 3,000 & $2.2 \times 10^{37}$ & 0.016    \\
Vela           & 0.089 & 11,200  &   250 & $6.3 \times 10^{36}$ & 0.008    \\
PSR B1951+32   & 0.039 & 107,000 & 2,400 & $3.7 \times 10^{36}$ & 0.03     \\ 
PSR B1706--44  & 0.102 & 17,400  & 1,800 & $3.4 \times 10^{36}$ & 0.09     \\
PSR B1046--58  & 0.124 & 20,400  & 3,000 & $3.1 \times 10^{36}$ & 0.14     \\
Geminga        & 0.237 & 340,000 &   160 & $3.3 \times 10^{34}$ & 0.20     \\
PSR B1055--52  & 0.197 & 540,000 & 1,500 & $3.0 \times 10^{34}$ & $\sim 1$ \\
\tableline
\tableline
\end{tabular}
\end{table}

\section{X-ray Pulsar and Nebula} 

The sharp main pulse with fast rise time in Figure 2 is indicative 
of nonthermal magnetospheric emission.
The weaker interpulse is equally prominent at soft and
hard energies.  That the
soft and hard X-ray pulse shapes of \psr\ are
the same within errors can be interpreted as
absence of evidence
for a separate component of surface thermal emission.
Although the \asca\ GIS source is
dominated by an unknown amount of extended emission (Halpern et al.
\begin{wrapfigure}{r}{2.75in}
\begin{center}
\psfig{figure=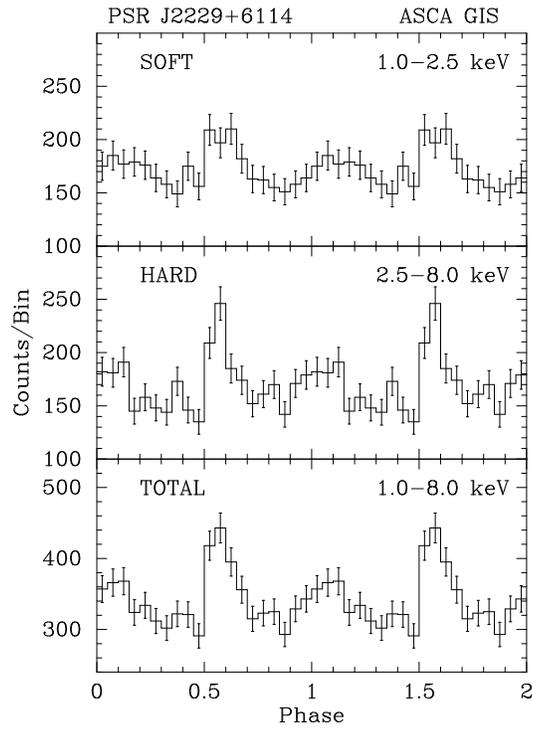,height=3.75truein}
\caption{X-ray pulse profile of PSR~J2229+6114 from the
{\it ASCA\/} GIS.  The instrumental time resolution is comparable
to the width of one phase bin. Unpulsed flux is dominated
by nebular emission.}
\end{center}
\end{wrapfigure}
2001a),
the subsequent observation by the \chandra\ ACIS-I revealed that
approximately 70\% of the 2--10 keV X-ray emission
comes from a centrally
peaked, diffuse nebula of radius $100^{\prime\prime}$ (Halpern et al. 2001b).
The diffuse X-rays are apparently confined within the
radio shell of similar radius discussed in the next section. 
The nebula has a power-law spectrum of photon index $\Gamma = 1.45 \pm 0.19$.
The pulsar itself has a marginally harder spectrum with photon
index $\Gamma = 0.99 \pm 0.27$, and unabsorbed 2--10 keV flux
$4.9 \times 10^{-13}$ erg~cm$^{-2}$~s$^{-1}$.
These fits are consistent with the
\asca\ measured $\Gamma = 1.51 \pm 0.14$ for the entire blended structure,
and the \asca\ value of $N_{\rm H} = (6.3 \pm 1.3) \times
10^{21}$ cm$^{-2}$ remains the most precise determination of the
column density to the pulsar.  We use this value of $N_{\rm H}$ 
to estimate the distance to the pulsar as $3 \pm 1$~kpc.  The distance
remains the most important, poorly determined property of \psr.
In addition to our X-ray estimate of 3~kpc, values as large as 12~kpc from the
radio pulsar dispersion measure of 200~cm$^{-3}$~pc
and the Taylor \& Cordes (1993)
free electron model, to as small as 800~pc from velocity
structure in surrounding H~I emission
(Kothes, Uyaniker, \& Pineault 2001; Kothes et al., these proceedings),
have been suggested.  

Although the short \chandra\ exposure barely reveals the complex
structures in the \psr\ nebula (Figure 3),
there is evidently an incomplete 
elliptical arc, similar to the structures that dominate the appearance
of the prototype of its class, the Vela PWN
(Helfand, Gotthelf, \& Halpern 2001), and a possible jet seen
as a point source $14^{\prime\prime}$ to the west of the pulsar.  
\section{Radio Nebula}

\begin{wrapfigure}{r}{2.75in}
\begin{center}
\psfig{figure=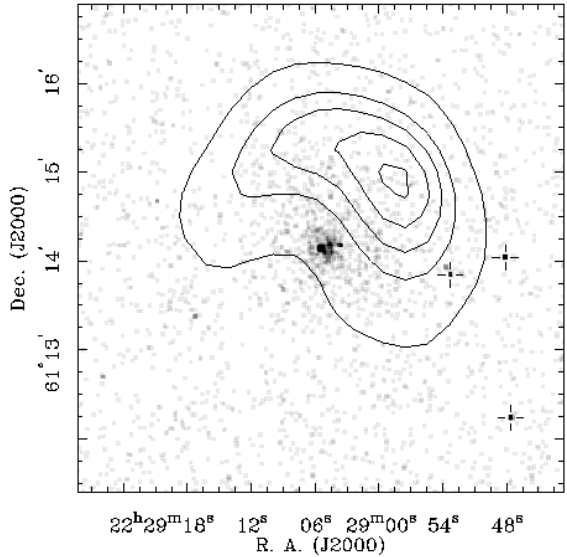,height=2.9truein,angle=270}
\caption{A portion of the {\it Chandra} ACIS--I image
showing \psr\ and its associated PWN ({\it greyscale}).  The
image is binned into $\approx 1^{\prime\prime}$ pixels and smoothed
with a 3-pixel top-hat filter.  The NVSS 20~cm map of
the partial shell \snr\ is overlaid ({\it contours}).}
\end{center}
\end{wrapfigure}

The incomplete radio shell that surrounds
\psr\ is unique in having an extremely
flat spectrum, $\alpha_r \approx 0.0$,
even though it has shell morphology (see Halpern et al. 2001a
for details).
It is clearly a non-thermal
source, as it is polarized at a level of $\approx 25\%$,
but its flat spectrum is not easily understood.
We assigned the name \snr\ to this radio structure.
Since the X-ray emission
appears to be largely confined within the radio shell, and since the
shell is too small ($r \approx 1.5$~pc) to be the blast wave of
a $10^4$ yr old supernova remnant, we conclude that the radio
emission comes from a shock driven into the surrounding medium either
by the motion of the pulsar or by the expansion of the PWN.  In the
bow-shock interpretation, we can relate the spin-down power,
$\dot E = 2.2 \times 10^{37}$ erg~s$^{-1}$ assumed to be carried
almost entirely by the PWN, to the velocity of the pulsar $v_p$,
the ambient density $n_{\rm H}$, and the radius of the shock $r_0$ via
$$\dot E\ =\ 4\pi\,r_0^2\,c\,\rho_0\,v_p^2\ =\  2.2\times 10^{37}
\ \left ({n_{\rm H} \over 0.01}\right )
\left ({d \over 3\,{\rm kpc}}\right )^2  \left ({v_p \over 90\,{\rm km\,s^{-1}}}
\right )^2\ \ {\rm erg\  s^{-1}}.$$
\noindent This would require a low-density medium, as might be appropriate at
a $z$-height of 150~pc or in a cavity previously evacuated by a
supernova explosion.  Alternatively, confinement of the radio nebula
by the {\it static\/} pressure of the surround ISM is possible 
in a higher-density medium if
$$\dot E\ =\ 4\pi\,r_0^2\,c\,(n_e+n_i)\,k\,T\ =\ 
2.2\times 10^{37}\ \left ({n_{\rm H} \over 0.9}\right )
\left ({d \over 3\,{\rm kpc}}\right )^2 
\left ({T \over 10^4\,{\rm K}}\right )\ \ {\rm erg\ s^{-1}}.$$

In addition to this compact radio nebula, Pineault \& Joncas (2000) discovered
the larger radio continuum source G106.3+2.7,
approximately $0.\!^{\circ}5 \times 1^{\circ}$ in extent,
which they classified as a supernova remnant.
Kothes et al. (2001) regard this structure,
which borders on the compact radio nebula \snr, as the remnant of the
supernova that gave birth to \psr.

\section{Comparing the PWN of PSR J2229+6114 to Vela's}

Morphological similarities between the \psr\ and Vela PWNe are the
incomplete elliptical arc, which Helfand et al. (2001) interpreted
as a cylindrical equatorial shock in the Vela pulsar's wind, and a 
possible jet, located on the minor axis of the ellipse,
assumed to be the projected rotation axis of the pulsar
(P.A. $280^{\circ}$ in Figure 3).  In the context of the Kennel \& Coroniti
(1984) model, the semi-major axis of the equatorial arc
can be interpreted as $r_s$, the radius of the MHD wind shock, while the
radio shell of \snr, which is either a bow-shock or a static bubble,
coincides with the outer extent
of the X-ray synchrotron nebula and denotes
$r_n$.  The Vela X-ray PWN is apparently confined
by the thermal pressure of its surrounding hot SNR, which
gives its X-ray emission a clear discontinuity.
In the case of the Crab $r_n/r_s \approx 20$, while for
\psr, $r_n/r_s \approx 9$, and Vela has $r_n/r_s \approx 2$.
Unlike the Crab Nebula, which is a fairly 
effective ``calorimeter'' of the pulsar spin-down power,
both \psr\ and Vela are characterized by extremely
inefficient X-ray emission, $L_x/\dot E \approx 8 \times 10^{-5}$
including pulsar and nebula.
This could be an indicator
of a highly magnetized pulsar wind that prevents an internal MHD 
shock, at the location of the X-ray arcs, from strongly compressing the flow.
Consequently, both the rapid post-shock outflow and the weak post-shock
magnetic field prevent the bulk of the wind energy from being radiated
away.

In the case of Vela, we have modelled the surface brightness of
its X-ray arcs in terms of a standing shock in a relativistic outflow.
If we assume that the arcs have cylindrical symmetry, velocities
of at least $0.5-0.7c$ are required to explain their asymmetric
brightness as Doppler boosting in the approaching side of the
flow, and dimming of the receding side.  If so, the entire postshock flow
from $r_s$ to $r_n \approx 2r_s$ must be relativistic, and the magnetization
parameter $\sigma$, defined as the ratio of Poynting flux to particle
flux in the pre-shock pulsar wind, must be of order 0.1 or greater
(see Figure 3 of Kennel \& Coroniti 1984).  In the Crab, of course,
the same model yields the small value $\sigma \approx 0.003$, which
is consistent with its large value of $r_n/r_s$ and small outflow velocity
$v<0.01c$ at $r_n$, the outer boundary of the nebula.

While there is much less detailed information about \psr\ than either
the Crab or Vela, inspection of Figure 3 gives the impression of
a significant asymmetry in the brightness of the arc,
therefore a high velocity at $r_s$.  A deeper \chandra\ image will
be obtained to quantify this effect, but from all appearances,
\psr\ resembles Vela more than it does the Crab.  Its surface
brightness declines severely with radius, which may be
the result of a post-shock magnetic field that declines as $1/r$
in the limit of large $\sigma$.  Whereas the values of 
$r_s$ in Vela ($1 \times 10^{17}$ cm)
and \psr\ ($5 \times 10^{17}$ cm) scale roughly as the expected 
$\dot E^{1/2}$, we note that $r_n \approx 1.5$~pc in \psr\ is a very large
value that probably indicates a lower ISM pressure
($ <3 \times 10^{-12}$ dyne~cm$^{-2}$) than in the case of the
smaller Vela nebula, which is confined by the pressure of its
hot surrounding SNR, estimated as $8.5 \times 10^{-10}$ dyne~cm$^{-2}$
(Markwardt \& \"Ogelman 1997; Helfand et al. 2001).
Here we have assumed $d = 250$ pc for Vela and $d = 3$ kpc for \psr.
However, a more complete understanding of the physics of \psr\ and its PWN
awaits a reliable determination of its distance, for which estimates 
ranging from 0.8 to 12 kpc have been suggested, an unacceptably large
disagreement.

\section{Broad-Band Spectrum of PSR J2229+6114}

The $\gamma$-ray spectrum of \source\ is parameterized as a
power-law of 
photon index $\Gamma = 2.24 \pm 0.14$
(Hartman et al. 1999), steep compared to all
\begin{wrapfigure}{r}{2.75in}
\begin{center}
\psfig{figure=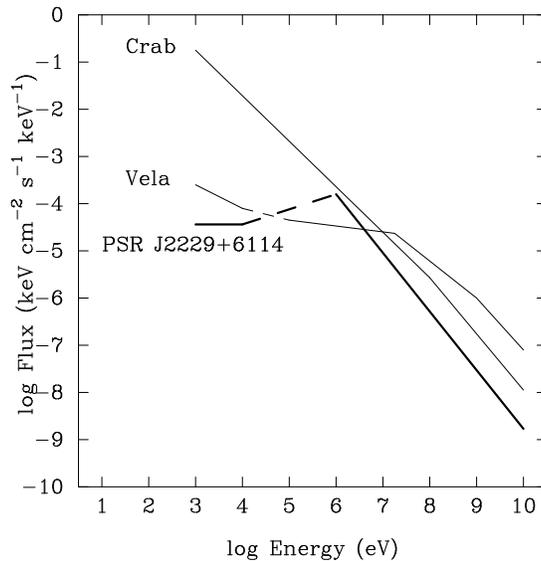,height=2.9truein,angle=0.}
\caption{The non-thermal, pulsed components of the
three most energetic $\gamma$-ray pulsars.\hfil}
\end{center}
\end{wrapfigure}
other EGRET pulsars
except the Crab, for which $\Gamma = 2.19 \pm 0.02$.
Iyudin et al. (1997) reported a source in the 0.75--3~MeV
band with COMPTEL, coincident with \source\ but with a much larger error box.
This detection is consistent in flux with an extrapolation of the
EGRET spectrum, but it exceeds an extrapolation of the 2--10 keV
spectrum of \psr\ to 1~MeV.  In Figure~4 we show a schematic representation
of the high-energy spectrum of \psr\ under the assumption that it is 
responsible for the coincident EGRET and COMPTEL sources.  In comparison
with the Crab and Vela, the only other pulsars detected by COMPTEL,
we see that \psr\ may be one of the brightest in the sky at 1~MeV,
even as it is relatively inconspicuous at radio through X-ray wavelengths.
This variety of broad-band spectra is encouraging of the hypothesis
that new rotation-powered pulsars will be discovered to be responsible for
more of the unidentified Galactic EGRET sources.

\end{document}